\documentclass[a4paper,fleqn]{cas-dc}

\usepackage[authoryear,longnamesfirst]{natbib}

\pdfoutput=1 

\def\tsc#1{\csdef{#1}{\textsc{\lowercase{#1}}\xspace}}
\tsc{WGM}
\tsc{QE}

\begin{document}
\let\WriteBookmarks\relax
\def\floatpagepagefraction{1}
\def\textpagefraction{.001}

\shorttitle{A Hybrid Data-driven Model of Ship Roll}    

\shortauthors{K. E. Marlantes, K. J. Maki}  

\title [mode = title]{A Hybrid Data-driven Model of Ship Roll}  

\nonumnote{The authors gratefully acknowledge ENGYS for the use of HELYX-OpenFoam.} 

\author{Kyle E. Marlantes}[orcid=0000-0002-0003-3617]
\cormark[1]
\ead{kylemarl@umich.edu}
\credit{Conceptualization, Investigation, Formal Analysis, Software, Data curation, Visualization, Writing - original draft}

\affiliation{organization={University of Michigan, Department of Naval Architecture and Marine Engineering},
            city={Ann Arbor},
            state={MI},
            country={USA}}

\author{Kevin J. Maki}
\credit{Supervision, Project administration, Conceptualization, Methodology, Funding acquisition, Writing - review \& editing}

\cortext[1]{Corresponding author}

\begin{abstract}
A hybrid data-driven method, which combines low-fidelity physics with machine learning (ML) to model nonlinear forces and moments at a reduced computational cost, is applied to predict the roll motions of an appended ONR Tumblehome (ONRT) hull in waves. The method is trained using CFD data of unforced roll decay time series--a common data set used in parameter identification for ship roll damping and restoring moments. The trained model is then used to predict wave excited roll responses in a range of wave frequencies and the results are compared to CFD validation data. The predictions show that the method improves predictions of roll responses, especially near the natural frequency.
\end{abstract}



\maketitle

\section{Introduction}\label{sect:intro}

Roll motion impacts the safety and operability of ships in ocean waves. The motion can sometimes be violent, leading to loss of cargo, crew injury, structural failure, or in some cases, loss of stability and capsize. However, accurately predicting roll motion in waves remains difficult due to strong nonlinear and viscous effects \citep{falzarano2015}, \citep{copuroglu2023}. The advent of the IMO Second Generation Intact Stability (SGIS) criterion, which bring dynamic failures, such as synchronous roll, parametric roll, and dead ship condition into the design evaluation phase, greatly increase the need for accurate--and practical--models of roll motion \citep{marlantes2021a}. Furthermore, the ever increasing demand to improve the economy, safety, and operability of ships, means that roll motion predictions remain a topic of considerable importance.\par

Understanding and modeling roll motion has captured the interest of many researchers for over a century. Experimental and full-scale observations underpin the earliest studies into the physics involved in the rolling motion of a ship \citep{froude1861}. The development of analytical models quickly followed observations, with work by \cite{dalzell1978}, \cite{roberts1985}, \cite{taylan2000}, \cite{bigalabo2022}, \cite{yu2022}, \cite{matsui2023}, and many others, demonstrating the strong nonlinearity present in the roll restoring and damping moments and the inadequacy of linear methods to provide accurate predictions \citep{nayfeh1989}, \citep{spyrou2000}, \citep{cotton2001}, \citep{kianejad2020}. The importance of nonlinear restoring and damping moments to roll motion is now widely recognized, with studies like \cite{wawrzynski2023} demonstrating the complex dynamics and the affect this has on the onset of capsize \citep{soliman1991}, \citep{falzarano1992}, \citep{lin1995}, parametric roll \citep{copuroglu2023}, and general seakeeping assessments \citep{ulmulk1994}, \citep{matsui2023}.\par

As a result, various forms of nonlinear hydrostatic restoring and damping moments have been posed. In general, quadratic damping or cubic damping, and cubic or quintic restoring are identified as valid models, though other functional relationships or piece-wise models have also been suggested, especially for large responses \citep{bassler2010}. However, under the assumption of an analytical form, it becomes necessary to identify the correct coefficients for a particular ship geometry \citep{kianejad2020}. For this reason, a significant number of researchers have presented parameter identification techniques to determine nonlinear restoring and damping coefficients from data. Examples include stochastic methods \citep{roberts1994}, energy methods, log-decrement methods \citep{roberts1985}, and others \citep{chan1995}, \citep{jang2010}, \citep{kim2015}, \citep{wassermann2016}, \citep{sun2021}, \citep{zhang2023}. Parameter identification methods have been applied to experimental and numerical data, typically at model scale.\par

Several different experimental approaches have been proposed to obtain suitable data, including 2D and 3D experiments in regular waves \citep{rodriguez2020}, irregular waves, roll decay tests, and harmonically excited roll motion tests (HERM) \citep{handschel2014}, \citep{wassermann2016}, \citep{olivaremola2018}, and many of these methods have also been implemented as numerical simulations \citep{begovic2015}, \citep{hashimoto2019}. To this end, scaling effects have been considered by some authors \citep{irkal2016}, \citep{kianejad2018}.\par

Despite the large body of work suggesting improved methods for parameter identification of roll restoring and damping coefficients, difficulties remain, especially related to the effects of appendages, such as bilge keels, and the complex nonlinear behavior exhibited by roll damping moments \citep{aloisio2006}, \citep{avalos2014}. Experimental and numerical investigations show that the damping moment is dominated by viscous effects \citep{zhang2019}, \citep{copuroglu2023}, especially flow separation and the dissipation of energy due to vortex shedding, which are not captured by potential flow methods \citep{jiang2020}. In recent years, Computational Fluid Dynamics (CFD) shows great promise to capture viscous effects in numerical simulations, and a large number of studies have been performed using CFD, including RANS and LES \citep{wilson2006}, \citep{gokce2018}, \citep{mancini2018}, \citep{kianejad2019}, \citep{irkal2019}, many demonstrating good comparison with model test data. However, the high cost of CFD precludes its use as a design tool for the evaluation of roll motion in a large number of operating conditions. For this reason, parameter identification still remains the most practical option.\par

In nearly all studies which are focused on parameter identification, the goal is to determine the parameters subject to a predefined form. However, some researchers have pointed out that the typical linear-plus-quadratic and linear-plus-cubic forms may not be adequate \citep{eissa2003}, \citep{wassermann2016}. For example, \cite{korpus1997} show certain components are in phase with the position and acceleration, not just the velocity. \cite{spyrou2008} and \cite{rajaraman2023} also suggest that fractional derivatives may play a role, further exposing the considerable complexity necessary to define an accurate model of ship roll motion.\par

Conclusions from decades of research suggest that there still remain two major challenges in predicting roll motions: 

\begin{enumerate}
\item classical quadratic, cubic, or quintic damping and restoring models do not capture the full nonlinear relationship exhibited by experimental or CFD results,
\item numerical methods, such as CFD, are far too expensive to be utilized as design tools for comprehensive assessment of roll motions.
\end{enumerate}

As a result, it is necessary to devise new methodology to model ship roll motion which works to address the two limitations. In this paper, the neural-corrector method of \cite{marlantes2022} is extended to model roll motions. The method is similar to classical parameter identification methods in that it learns the form of the nonlinear restoring, damping, and added mass moments from data, however, it makes no assumptions about the functional form of the moments. Furthermore, the model can be evaluated at a cost similar to classical ODE methods, with only a minor additional cost due to the evaluation of small embedded neural networks. \par

The proposed method is a hybrid machine learning (ML) method because it combines low-order physics with data-driven techniques. Hybrid methods have gained considerable attention in recent years as they show promise to overcome many of the challenges associated with physics-only or data-only methods \citep{willard2020}. One of the earliest examples of such methods applied to ship hydromechanics is given by \cite{weymouth2014}, and the idea has since been expanded to ship motions \citep{wan2018}, \citep{marlantes2021b}, \citep{schirmann2022}, \citep{schirmann2023}, maneuvering \citep{skulstad2021}, resistance \citep{yang2022}, and recently hull-form optimization \citep{bagazinski2023}, and other areas. The main benefit that hybrid methods offer over physics-only methods is the inference cost is typically much lower than a high-fidelity physics-only method, yet the fidelity of the predictions remain improved. The benefit over data-only methods is that the training data requirements tend to be much lower because the embedded physics introduce inherent knowledge to the system. This also means that hybrid methods are more transferable, meaning they perform better when making predictions in conditions that are different from the training data set \citep{yang2022}. Transferability is a critical requirement for the present application in marine dynamics as evaluation in differing wave conditions, forward speeds, and loading conditions can lead to a large number of conditions and training data cannot be sampled from each expected condition.\par

A few studies should be specifically recognized for their similarity to the present work even though they approach the problem in a different manner. \cite{chen2019} applied data-only ML techniques to the roll parameter identification problem. While the form of the restoring and damping is still quadratic, it is one of few examples of ML being applied to the roll parameter identification problem. \cite{somayajula2017} propose a novel system identification technique which is used to identify an improved roll damping model from responses in irregular waves. The method is similar in that it takes a correction approach to the low-order linear physics. And lastly, the work by \cite{jang2010} considers the inverse problem using Tikhonov regularization, providing another example of work which attempts to not make an assumption about the form of the damping model.\par

This paper is organized into five remaining sections. Section \ref{sect:analytic} will introduce a simplified analytical model to  highlight the difficulties associated with modeling ship roll motions. Section \ref{sect:method} will outline the proposed methodology in overcoming the challenges discussed. Section \ref{sect:cfd} gives a description of a CFD case of the ONR (Office of Naval Research) Tumblehome hull at model scale which will be utilized to generate training and validation data. Section \ref{sect:result} gives the results of the proposed method applied to the CFD data and, lastly, Section \ref{sect:concl} gives conclusions.

\section{Analytical Model}\label{sect:analytic}

To better understand the challenges associated with predicting ship roll motion, it is helpful to first consider a simplified analytical model. A single-degree-of-freedom (1-DOF) Duffing equation, as given by Eq. \eqref{duffeq}, is a common low-order model of ship roll. Here we use the form of the equation studied by \cite{amaki2022}, where $\phi$ is the state variable roll angle, $G(\phi)$ is the restoring moment, $\alpha$, $\beta$, and $\gamma$ are roll damping coefficients, $\omega_{0}$ is the natural frequency of the ship in rad$\cdot$s$^{-1}$, $M(t)$ is the wave excitation moment, $GZ$ is the righting arm, and $GM$ is the upright transverse metacentric height. A direct numerical solution to Eq. \eqref{duffeq} is certainly not difficult, however, the equation offers fundamental insight into the nonlinearity present in ship roll responses.

\begin{align}
\ddot{\phi} &+ \alpha \dot{\phi} + \beta \dot{\phi} | \dot{\phi} | + \gamma \dot{\phi}^{3} + G(\phi) = M(t) \label{duffeq} \\
G(\phi) &= \sum_{i=1}^{5} G_{i} \phi^{i}, G_{1} = \omega_{0}^{2}, G_{i} = \omega_{0}^{2} \frac{GZ_{i}}{GM} \label{stiffnl}
\end{align}

The excitation moment due to a regular incident wave is computed using the wave slope according to Eq. \eqref{extmom}, where $\Gamma$ is the effective wave slope coefficient, $k_{w}$ is the wave number, $\zeta_{w}$ is the wave amplitude, $\omega$ is the wave frequency in rad$\cdot$s$^{-1}$, and $\epsilon$ is the phase angle. Also given by Eq. \eqref{waveel} is the wave elevation $\eta(t)$ of the incident wave.\par

\begin{align}
M(t) &= \Gamma \omega_{0}^{2} k_{w} \zeta_{w} \cos{(\omega t + \epsilon )} \label{extmom} \\
\eta (t) &= \zeta_{w} \sin{ ( \omega t + \epsilon ) } \label{waveel}
\end{align}

\subsection{Restoring Moment}

\begin{figure}
	\centering
		\includegraphics{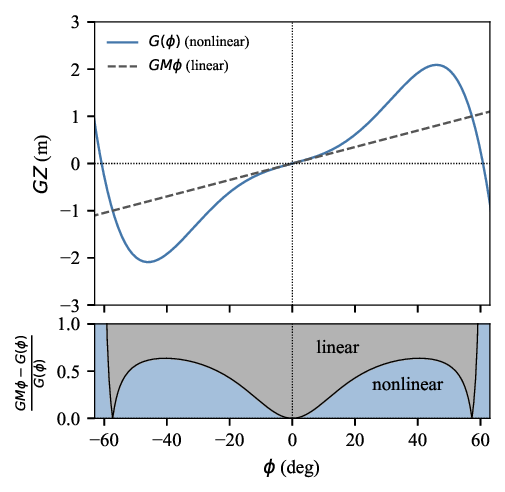}
	  \caption{Nonlinear $GZ$ curve used in Duffing equation to model roll response. The lower panel shows the ratio of the nonlinear part of the curve $G(\phi)-GM\phi$ and the linear approximation $GM\phi$ to the total $GZ$.}\label{duffgz}
\end{figure}

The primary characteristic of the Duffing equation is the restoring moment $G(\phi)$ is made nonlinear by including polynomial terms of order greater than one, as given by Eq. \eqref{stiffnl}. Because the restoring moment of a ship in roll is governed by its righting moment and associated $GZ$ curve, a polynomial form of 3rd or 5th order is a reasonable approximation of the nonlinear variation in the moment. This nonlinear variation is a result of the hull form, which is typically not equisymmetric about the axis of inclination. Figure \ref{duffgz} shows an example of a $GZ$ curve, where 5th order behavior can be observed. Also shown is the corresponding linear model, which is simply a straight line with a slope equal to the $GM$.\par

Figure \ref{duffgz} shows that for roll angles larger than about 10 degrees, the nonlinear components of the restoring moment contribute greater than 25\% of the total restoring moment. As a result, the true restoring moment will deviate considerably from the linear model. This is one primary reason why roll motions are strongly nonlinear at large amplitudes. In a purely linear system, the stiffening and softening of the restoring moment are not present. The linearization of the restoring moment fundamentally changes the dynamics of the model. Primarily, a nonlinear restoring moment will result in a shifting natural frequency \citep{wawrzynski2016}, \citep{bigalabo2022}, super- and sub-harmonics \citep{cardo1981}, hysteresis \citep{ulmulk1994}, and chaotic dynamics \citep{falzarano1992}, \citep{lin1995}, \citep{cotton2001}.\par

One important consideration which is not captured by the 1-DOF Duffing equation, even with a 5th order restoring moment, is the effects of coupling with other modes of motion. Several authors have identified the importance of additional DOF for accurate roll predictions \citep{ulmulk1994}, \citep{mancini2018}, \citep{hashimoto2019}. In principle, this means that the roll restoring moment cannot be modeled as a function of only roll angle, but must also include terms related to heave and pitch. The linearized coupling forces and moments are established and well-understood, but an analytical relationship of the nonlinear restoring forces and moments does not exist. Typically, the effects are considered using direct pressure integration in a high-fidelity numerical method.\par

\subsection{Damping Moment}

In a similar manner, nonlinear damping is included in Eq. \eqref{duffeq} and parameterized by the coefficients $\alpha$, $\beta$, and $\gamma$. Cubic and quadratic damping terms are common analytical approximations of the form of the damping moment when modeling nonlinear roll responses. However, studies show that depending on the amplitude and frequency, the damping can be largely concentrated into one or all of the terms, and as with the restoring moment, the nonlinear damping terms grow much faster when roll velocity is large. Therefore, when large roll amplitudes typically coincide with large roll velocities, and nonlinear  terms govern the damping moment.\par

If we consider Eq. \eqref{duffeq} without the nonlinear terms, i.e. $G_{i}=0$, $i>1$ and $\beta$=$\gamma$=0, we can express the complex roll response amplitude $\bar{\phi}$ as Eq. \eqref{rollRAO}, where $\bar{M}$ is the complex wave excitation amplitude which oscillates at a single angular frequency $\omega$, and $i$ is the imaginary number.

\begin{equation}
\bar{\phi} = \frac{\bar{M}}{-\omega^{2} + \alpha\omega i + G_{1}} \label{rollRAO}
\end{equation}

Now, if the excitation frequency $\omega$ is taken equivalent to the roll natural frequency $\omega_{0}=\sqrt{G_{1}}=\omega$ then Eq. \eqref{rollRAO} reduces to Eq. \eqref{rollampl}, where $|\cdot|$ indicates the amplitude of the complex argument.

\begin{equation}
|\bar{\phi}| = \frac{|\bar{M}|}{\alpha\omega_{0}} \label{rollampl}
\end{equation}

Eq. \eqref{rollampl} shows that the amplitude of the roll response is entirely dependent on the damping coefficient $\alpha$ when resonance occurs $\omega=\omega_{0}$. Therefore, without an accurate damping model, the obtained predictions of $|\bar{\phi}|$ will be incorrect.\par

Classically, roll damping is divided phenomenologically into components, such as wave radiation damping, skin friction damping, eddy damping, lift damping, and appendage damping, as introduced in the pioneering work of \cite{ikeda1978}. Potential flow methods, which assume inviscid and irrotational flow, compute only the wave radiation damping since it can be derived from the solution to the radiation problem. The radiation component is typically quite a small component of the total damping, which is why potential flow methods greatly over-predict roll responses near resonance without damping corrections. The remaining components are all due to viscous effects: skin friction is realized in the boundary layer and depends on surface roughness, eddy damping, which is often the largest damping component, is related to viscous separation around the hull as energy is lost due to the formation of vortex structures. Appendage damping is almost entirely dominated by viscous separation, as flow over bilge keels, skegs, and rudders leads to flow separation and energy dissipation. \par

At present, in industrial simulations, damping components are most often computed using Ikeda methods, \citep{ikeda1977}, \citep{ikeda1978}, \citep{ikeda1978}, \citep{himeno1981}, \citep{ikeda1983} which offer semi-empirical formulas and are relatively easy to implement alongside potential flow solvers. However, a number of researchers have identified limitations to the Ikeda methods \citep{haddara1994} and improvements are still being suggested to make the methods more applicable to a larger range of designs \citep{katayama2023}. As a result, roll damping remains one of the greatest challenges in ship motion predictions. \par

\subsection{Added Mass Moment of Inertia}

Eq. \eqref{duffeq} implicitly assumes linear added mass, and the equation is normalized by the total mass (including added mass), given that the coefficient of $\ddot{\phi}$ is one. However, added mass moment in roll is known to be nonlinear as well \citep{zhang2019}, \citep{bigalabo2022} and can have marked impacts on the natural frequency and the effects of appendages such as bilge keels \citep{kianejad2019b}, \citep{kianejad2020}, \citep{jiang2020}. While most studies linearize added mass, it is important to recognize that this may not always be an appropriate simplification.

\section{Proposed Method}\label{sect:method}

The proposed method is based on the neural-corrector method of \cite{marlantes2022}, which embeds a learned data-driven force correction in a low-order equation of motion. In previous work, the method is applied to predict the heave and pitch motions of a planing boat in head seas using a synthetic data set. In the present work, the neural-corrector method is extended to model roll motions, with specific application to modeling nonlinear restoring, damping, and added mass using CFD data.\par

The proposed method is developed using models of ship roll at two levels of fidelity, where the superscript $(l)$ indicates low-fidelity and $(h)$ indicates high-fidelity. Under this notation, Eq. \eqref{hfeom} is the high-fidelity model, where the solution $\phi^{(h)}$ is what we would like to obtain and is assumed to be computationally costly, i.e. derived from CFD or model testing. Eq. \eqref{lfeom} is the low-fidelity model, where the linear added mass $A_{44}^{0}$, potential wave damping $B_{44}^{0}$, and hydrostatic restoring $C_{44}^{0}$ coefficients are computed at equilibrium. $M_{w}^{(l)}$ is the low-order wave excitation moment.\par

\begin{align}
I_{xx}\ddot{\phi}^{(h)} &= M^{(h)} \label{hfeom} \\
(I_{xx}+A_{44}^{0})\ddot{\phi}^{(l)} + B_{44}^{0}\dot{\phi}^{(l)} + C_{44}^{0}\phi^{(l)} &= M_{w}^{(l)} \label{lfeom}
\end{align}

Eq. \eqref{nceom} gives the proposed hybrid equation of motion, where the superscript $*$ indicates the approximate high-fidelity state. The linearized physics are retained and a data-driven force correction, $\delta$ is included on the right-hand-side of the equation. The intent is to design $\delta$ such that the error between $\phi^{(h)}$ and $\phi^{*}$ is minimized. As shown in \cite{marlantes2022}, $\delta$ is a complex term, including high-order contributions from the restoring, damping, and added mass moments and so it cannot be efficiently modeled using analytical techniques.\par

\begin{equation}
(I_{xx}+A_{44}^{0})\ddot{\phi}^{*} + B_{44}^{0}\dot{\phi}^{*} + C_{44}^{0}\phi^{*} = M_{w}^{(l)}  + \delta(\phi^{*},\dot{\phi}^{*},\ddot{\phi}^{*}) \label{nceom}
\end{equation}

In this work, the low-fidelity wave excitation $M_{w}^{(l)}$ is derived using a long-wave assumption, as given by Eq. \eqref{wmom}. This choice is purely out of convenience, as the linear coefficients are already known. Other more accurate models can certainly be used in Eq. \eqref{nceom} without a loss of generality.

\begin{equation}
M_{w}^{(l)}= k_{w}\zeta_{w} \left( (C_{44}^{0} - A_{44}^{0}\omega^{2})\cos{(\omega t)} - B_{44}^{0}\omega \sin{(\omega t)} \right) \label{wmom}
\end{equation}

\begin{figure}
	\centering
		\includegraphics{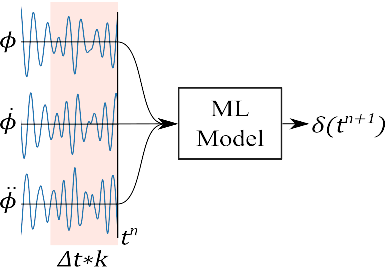}
	  \caption{Design of ML model for use in neural-corrector method.}\label{ncmethod}
\end{figure}

The design of $\delta$ is discussed in more detail in \cite{marlantes2023}, but it is important to highlight that the term maps state variables to forces or moments, as shown in Figure \ref{ncmethod}. The term can be formulated as a function of the low-fidelity state $\phi^{(l)}$ so that a low-fidelity update is required in addition to an update to $\phi^{*}$ at each time step. However, in this work $\delta$ is designed to be a function of $\phi^{*}$ directly, making the update process fully auto-regressive. Retaining the linear force terms in the equation of motion as analytical expressions and not learning them in $\delta$ also leads to improved solution stability in the presence of errors in $\delta$, as shown in \cite{marlantes2023}. \par

Eq. \eqref{nceom} is solved numerically to obtain $\phi^{*}$, $\dot{\phi}^{*}$, $\ddot{\phi}^{*}$. To accommodate numerical integration, the machine learning model for $\delta$ follows a many-to-one paradigm, where the input features are $k$-length sequences of previous state $\phi^{*}$, $\dot{\phi}^{*}$, $\ddot{\phi}^{*}$ and the output of the model is the moment correction $\delta$ at the next time step, as shown in Figure \ref{ncmethod}. In this way, the model is trained for a particular time step size. 

The proposed method does not require a specific ML architecture, though both Long Short-Term Memory (LSTM) and feed-forward, fully-connected neural networks have been used in prior work \citep{marlantes2022}, \citep{marlantes2023}. Because $k$ is typically less than 200 (spanning only 2~s for a timestep of 0.01~s) for typical hydrodynamic forces in marine dynamics problems, the need for memory as provided by LSTMs and Transformer networks is unnecessary and feed-forward networks with two or three layers are adequate. In this work, the ML model is a feed-forward, fully-connected neural network with 3 hidden layers, 10 nodes per layer, and ReLU activation functions.

\subsection{Roll Decay Time Series as Training Data}\label{sect:rolldecay}

To obtain training data for $\delta$, Eq. \eqref{nceom} is simply solved for $\delta$ and known high-fidelity $\phi^{(h)}$, $\dot{\phi}^{(h)}$, and $\ddot{\phi}^{(h)}$ time series are then used to compute a corresponding $\delta$ time series, as given by Eq. \eqref{delta}. It is therefore necessary to have an estimate of the linear coefficients for the hull under consideration, which can be obtained using a strip theory program. 

\begin{equation}
\delta = (I_{xx}+A_{44}^{0})\ddot{\phi}^{(h)} + B_{44}^{0}\dot{\phi}^{(h)} + C_{44}^{0}\phi^{(h)} \label{delta}
\end{equation}

Note that because the ML model is modeling forces or moments, and not the responses directly, the high-fidelity responses do not necessarily have to be the same as the expected predictions. In other words, responses computed in regular waves could be used for training data, and predictions could be made in irregular waves, for example. This is a great convenience of the proposed method. However, whatever time series are used must capture large enough amplitudes so that information about the nonlinear nature of the force or moment is embedded.\par

Since roll motion is the focus of the current work, it is natural to use unforced roll decay time series from a roll decay simulation or experiment as $\phi^{(h)}$ to derive $\delta$ and train the underlying ML model. Roll decay time series also intrinsically spans a range of roll amplitudes and identifies the damped natural frequency of the roll motion. Furthermore, the data are commonly available, either numerically or experimentally. 

\section{CFD Case of the ONR Tumblehome}\label{sect:cfd}

\begin{figure}
	\centering
		\includegraphics[width=0.5\textwidth]{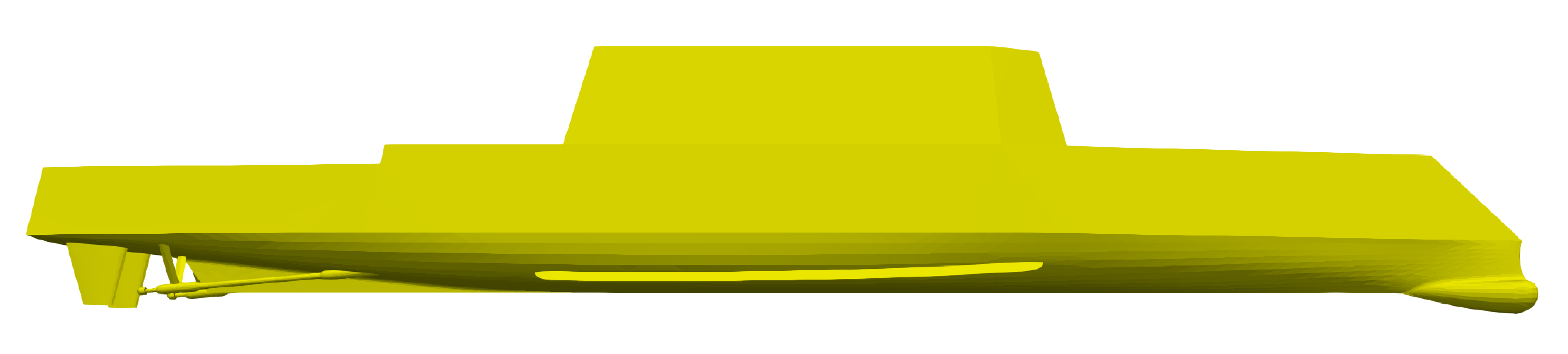}
	  \caption{A profile view of the ONRT hull geometry.}\label{onrtprof}
\end{figure}

To generate high-fidelity training and test data, a CFD case of a model scale (1:49) ONR Tumblehome (ONRT) hull is prepared using HELYX-OpenFOAM. The ONRT is a preliminary design of a modern surface combatant and is widely used for numerical and experimental ship hydrodynamics research. Figure \ref{onrtprof} shows the profile view of the hull, where the sonar dome, bilgekeels, propeller shafts and struts, skeg, and rudders are visible. Table \ref{onrt} gives the particulars of the ONRT model as used in this work \citep{bishop2005}. Also given are the model scale linear added mass, potential wave damping, and hydrostatic stiffness coefficients for the equilibrium condition, which are computed using a strip theory program.

\begin{table}
\caption{ONRT Model Particulars}\label{onrt}
\begin{tabular}{|c|c|c|c|}
\toprule

Scale (-) & 1:49 & \\
$L_{WL}$ (m) & 3.147 & Depth $D$ (m) & 0.266 \\
$B_{WL}$ (m) & 0.384 & Draft $T$ (m) & 0.112 \\
\midrule
Mass $m$ (kg) & 72.6 & $A_{44}^{0}$ (kg-m$^2$) & 0.4 \\
$I_{xx}$ (kg-m$^{2}$) & 1.5215 & $B_{44}^{0}$ (kg-m$^2$/s) & 0.3 \\
$V_{CG}$ (m, abl) & 0.112 & $C_{44}^{0}$ (kg-m$^2$/s$^2$) & 61.5 \\
\bottomrule
\end{tabular}
\end{table}

The CFD case uses the incompressible PIMPLE algorithm to solve the Reynolds Averaged Navier-Stokes Equations (RANSE) for a multi-phase air and water domain. A Volume-of-Fluid (VOF) approach is used to capture the interface between the air and water. Variable timestepping is allowed up to a maximum Courant number of 2.0, a maximum $\alpha$-Courant number of 0.5, or a maximum timestep of 0.001~s, whichever is smaller. 

\begin{figure}
	\centering
		\includegraphics[width=0.5\textwidth]{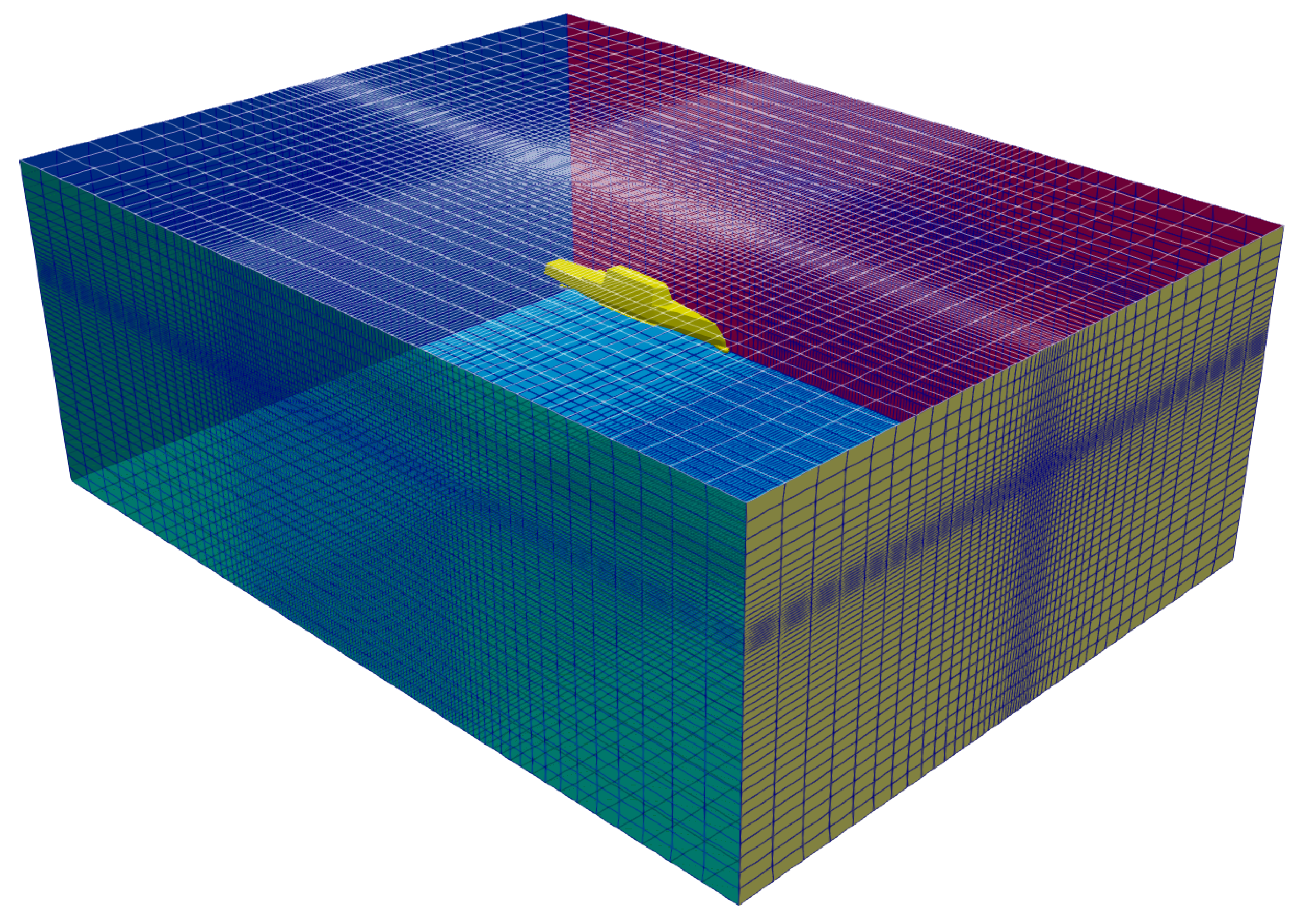}
	  \caption{The computational domain used in the CFD case for the ONRT. The faces are identified by color as follows: \textbf{\textcolor{Goldenrod}{front}}, \textbf{\textcolor{blue}{back}}, \textbf{\textcolor{ForestGreen}{starboard}}, \textbf{\textcolor{red}{port}}, \colorbox{black}{\textbf{\textcolor{white}{top}}}, \textbf{\textcolor{cyan}{bottom}}.}\label{onrtdom}
\end{figure}

Figure \ref{onrtdom} shows the computational domain, where the boundaries are colored for identification. The domain is 12.6~m long (4$L_{WL}$), 9.44~m wide (3$L_{WL}$), with water ($\alpha$=1) extending through the bottom 3.15~m ($L_{WL}$), and air ($\alpha$=0) in the top 1.51~m (0.5$L_{WL}$) following guidance in \cite{ittc2017b}. Notice the refinement in the mesh near the freesurface. Table \ref{cfdbcs} gives the boundary conditions specified on each boundary face. A $k\omega$-SST turbulence model is used, with wall functions at the hull surface. 

\begin{table*}
\caption{Boundary Conditions in ONRT CFD Case}\label{cfdbcs}
\begin{tabular*}{\textwidth}{|c|c|c|c|c|c|c|}
\toprule
Face & U & alpha.water & p\_gh & k & omega & nut \\
\midrule
\textbf{\textcolor{Goldenrod}{front}} & symmetry & symmetry & symmetry & symmetry & symmetry & symmetry \\
\textbf{\textcolor{blue}{back}} & symmetry & symmetry & symmetry & symmetry & symmetry & symmetry \\
\colorbox{black}{\textbf{\textcolor{white}{top}}} & \scriptsize \makecell{ press.InletOutletVelocity \\ uniform ( 0 0 0 )} & \scriptsize \makecell{inletOutlet \\ uniform 0} & \scriptsize \makecell{totalPressure \\ gamma 1.4 \\ p0 uniform 0} & \scriptsize \makecell{inletOutlet \\ uniform 0.0002 } & \scriptsize \makecell{inletOutlet \\ uniform 0.01}& \scriptsize \makecell{calculated \\ uniform 0.001}\\
\textbf{\textcolor{cyan}{bottom}} & \scriptsize waveVelocity & \scriptsize waveAlpha & \scriptsize zeroGradient & \scriptsize \makecell{fixedValue \\ uniform 0.0002 } & \scriptsize \makecell{fixedValue \\ uniform 0.01}& \scriptsize \makecell{calculated \\ uniform 1e-05}\\
\textbf{\textcolor{ForestGreen}{stbd}} & \scriptsize waveVelocity & \scriptsize waveAlpha & \scriptsize zeroGradient & \scriptsize \makecell{turb.Int.Kin.EnergyInlet \\ intensity 0.05 \\ uniform 0.1 } & \scriptsize \makecell{inletOutlet \\ uniform 0.01}& \scriptsize \makecell{calculated \\ uniform 0.001}\\
\textbf{\textcolor{red}{port}} & \scriptsize waveVelocity & \scriptsize waveAlpha & \scriptsize zeroGradient & \scriptsize \makecell{fixedValue \\ uniform 0.0002 } & \scriptsize \makecell{fixedValue \\ uniform 0.01}& \scriptsize \makecell{calculated \\ uniform 1e-05}\\
hull & \scriptsize \makecell{movingWallVelocity \\ uniform ( 0 0 0 )} & \scriptsize zeroGradient & \scriptsize \makecell{fixedFluxPressure \\ uniform 0} & \scriptsize \makecell{kqRWallFunc. \\ uniform 1e-20} & \scriptsize \makecell{omegaWallFunc. \\ uniform 1} & \scriptsize \makecell{nutUSpaldingWallFunc. \\ uniform 0.001} \\
\bottomrule
\end{tabular*}
\end{table*}

\begin{figure}
	\centering
		\includegraphics[width=0.5\textwidth]{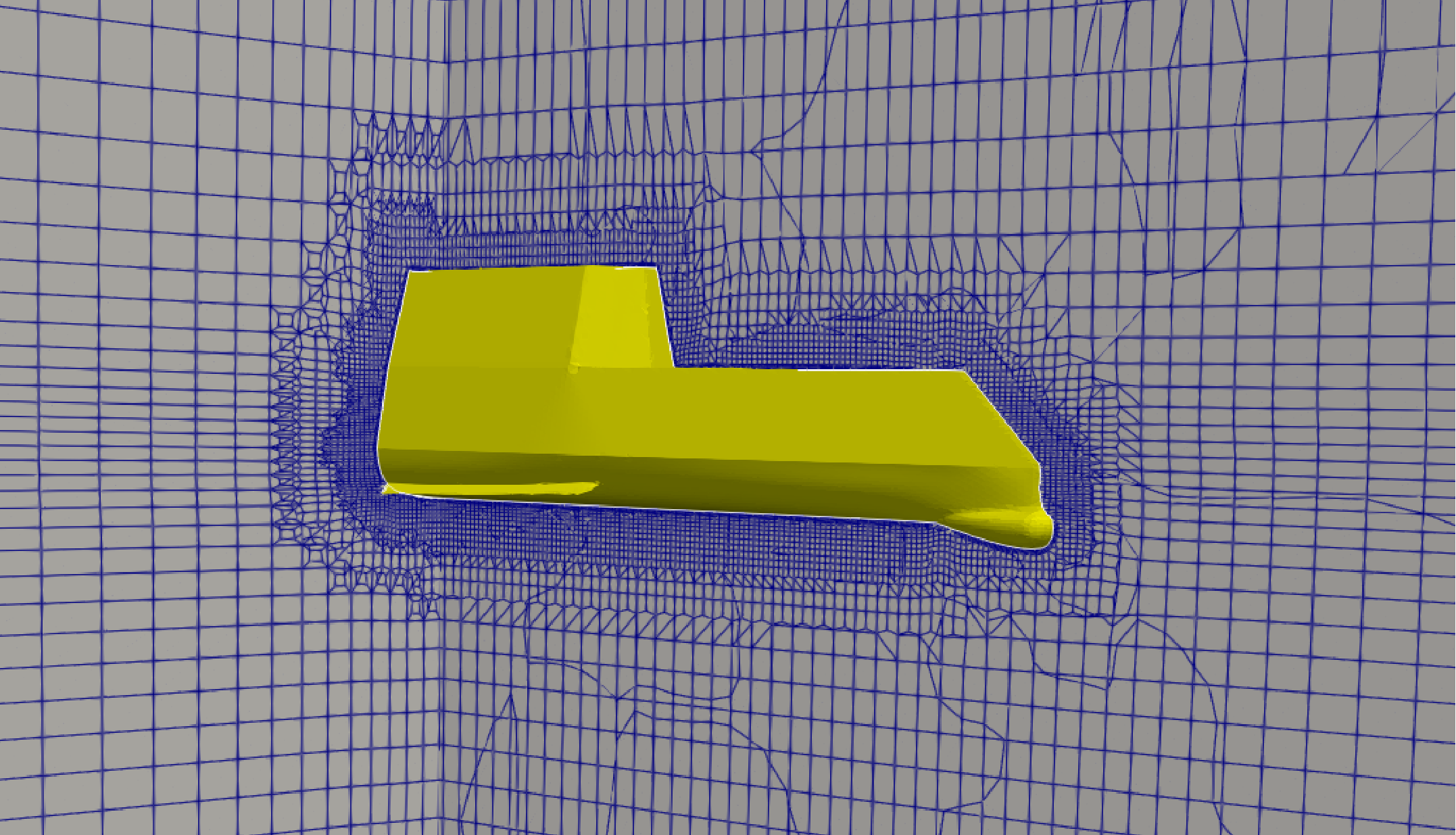}
	  \caption{A view of the unstructured hexahedral mesh used to simulate the roll response of the ONRT. The mesh is intentionally coarse, with a total count of 658,473 cells.}\label{cfdmesh}
\end{figure}

The mesh is generated using the HelyxHexMesh utility with five extruded layers on the hull surface. The mesh is intentionally coarse, with a total of 658,473 cells. Figure \ref{cfdmesh} shows the final mesh near the hull as used in the simulations. A customized dynamic mesh method allows the input of the initial inclination of the hull for the roll decay simulation with or without forward speed. The mesh is fixed to the hull, but the entire domain is free to move with the rigid-body motion of the hull which is driven by the fluid forces acting on the hull patches. To initialize the roll decay simulations, the dynamic mesh is ramped up to the desired inclination over a 1~s interval before the mesh is released.

\begin{figure}
	\centering
		\includegraphics{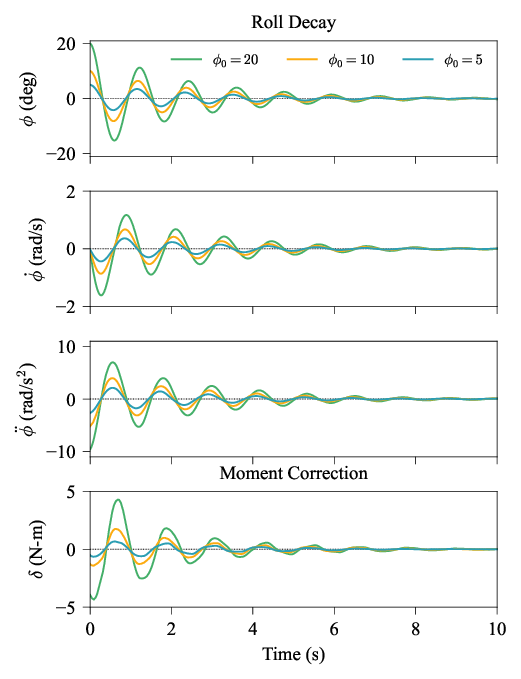}
	  \caption{Roll decay time series $\phi^{(h)}$, $\dot{\phi}^{(h)}$, $\ddot{\phi}^{(h)}$ of the ONRT at zero forward speed computed using the CFD case for use as training data in Eq. \eqref{delta}. Three different initial conditions are shown: $\phi_{0}$=5~deg, 10~deg, and 20~deg, however, only the $\phi_{0}$=20~deg case is used for training data.}\label{cfddecay}
\end{figure}

Figure \ref{cfddecay} shows roll decay time series obtained from the CFD case for three different initial conditions: inclinations of 5 degrees, 10 degrees, and 20 degrees. The total time series length for a single simulation is 20 seconds. Also shown in the figure are the resulting force correction time series, computed using Eq. \eqref{delta} and the linear coefficients given in Table~\ref{onrt}. The responses $\phi^{(h)}$, $\dot{\phi}^{(h)}$, $\ddot{\phi}^{(h)}$ and corresponding $\delta$ in Figure \ref{cfddecay} comprise the entire training data set available to train the ML model. However, training on all three time series proves to be less effective than training on only the $\phi_{0}$=20~deg case, likely due to the lack of nonlinear variation in the small amplitude cases. Therefore, the final training data set consists of only the $\phi_{0}$=20~deg decay simulation: a total of 20 seconds of time series with a downsampled time step of 0.001~seconds.

In addition to the roll decay simulations, roll responses in five different wave periods, 1.00~s, 1.14~s, 1.25~s, 1.39~s, and 2.00~s, all using a wave amplitude of 0.05~m are simulated using the same domain and model set up as the roll decay simulations. The responses are generated not as training data, but as validation data to test the accuracy of the proposed method. Regular waves are generated using the waves2Foam toolbox, which utilizes a relaxation zone technique to enforce the wave velocity and $\alpha$-field in the inlet and outlet regions \citep{jacobsen2012}. Two rectangular relaxation zones, one on the starboard side inlet, and one on the port side outlet, are specified, with the inboard edge of the relaxation zones extending to 1.0~m off the centerline of the domain, or about 2.25$B_{WL}$ from the side of the hull. A spatial-implicit relaxation scheme is used and the relaxation weights are specified using a third-order polynomial.\par

\section{Results}\label{sect:result}

\begin{figure}
	\centering
		\includegraphics{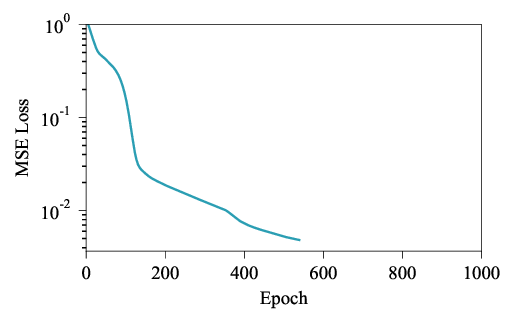}
	  \caption{Training loss for a feed-forward, fully-connected neural network with 3 hidden layers, 10 nodes per layer, reLU activation functions. The model learns the many-to-one relationship between $k$-length sequences of the state variables $\phi$, $\dot{\phi}$, $\ddot{\phi}$ at $t=n-k .. n$ and the force correction $\delta$ at $t=n+1$ for $k=5$.}\label{loss}
\end{figure}

The roll decay data for $\phi_{0}$=20~deg from Figure \ref{cfddecay} are used to train a feed-forward, fully connected neural network model as described in Section \ref{sect:method}. A stencil-length of $k$=5 is selected for the configuration, which corresponds to 0.005~seconds with a 0.001~s timestep. The model is trained using the Adam optimizer \citep{kingma2015} and a Mean-Squared-Error (MSE) loss function for up to 1000 epochs, stopping early when the loss does not improve greater than 1e-04 after ten consecutive iterations. Figure \ref{loss} shows the training loss for the model.

\begin{figure}
	\centering
		\includegraphics{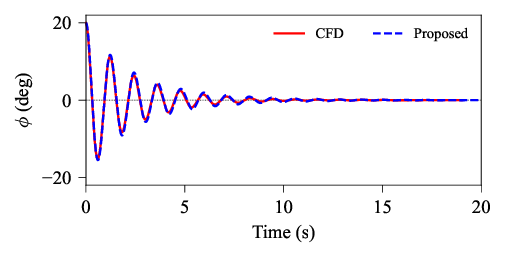}
	  \caption{Roll decay time series comparison between the CFD results $\phi^{(h)}$ and the proposed model $\phi^{*}$ of the ONRT at zero forward speed for initial condition $\phi_{0}$=20~deg.}\label{decaytest}
\end{figure}

Using the trained model, Eq. \eqref{nceom} is solved numerically using a first-order forward Euler scheme. To check that the learned force correction is indeed performing as expected within the integration scheme, Eq. \eqref{nceom} is first solved without external forcing (i.e. $M_{w}^{(l)}=0$) using the same initial conditions that are used in the CFD simulations in Section \ref{sect:cfd}. Figure \ref{decaytest} shows the solution for the initial condition $\phi_{0}$=20~deg. As expected, the proposed method closely follows the high-fidelity CFD decay time series.\par

\begin{figure}
	\centering
		\includegraphics{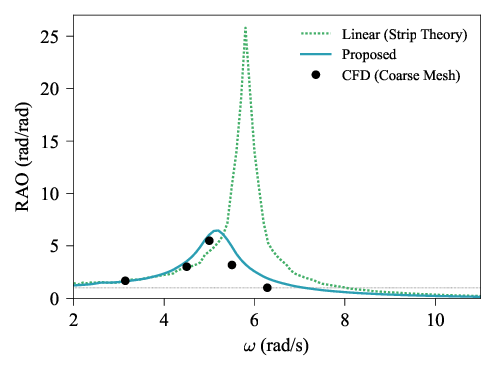}
	  \caption{Predicted forced roll response RAO for the ONRT at zero forward speed in regular waves over a range of wave excitation frequencies.}\label{result}
\end{figure}

Reproducing the training data time series is encouraging, however, the proposed method is not useful if it cannot be used in conditions that differ from the training data set. Furthermore, the method must perform well in the presence of wave excitation moments, since the primary objective of this work is to develop a model which can be used to efficiently assess the roll motion of a ship in waves. To test this, the wave excitation moment $M_{w}^{(l)}$ is modeled using Eq. \eqref{wmom} using the linear coefficients from Table \ref{onrt}. Eq. \eqref{nceom} is solved for wave frequencies ranging from 2.0 to 12.0~rad$\cdot$s$^{-1}$ with a constant wave amplitude of 0.05~m and at-rest initial conditions of $\phi_{0}$=$\dot{\phi}_{0}$=$\ddot{\phi}_{0}$=0. The steady-state peak responses from each time domain solution are used to compute the RAO of the response, normalized by the  wave slope $k_{w}\zeta_{w}$. The resulting RAO is shown in Figure \ref{result} alongside the regular wave results that are computed using the CFD case. Also shown is the linear RAO, computed using strip theory, for the same loading condition. \par

\begin{figure}
	\centering
		\includegraphics{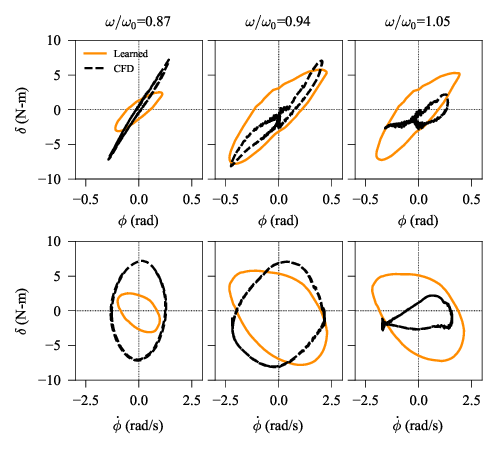}
	  \caption{Lissajous curves for the learned and CFD-derived $\delta$ vs $\phi$, $\dot{\phi}$, and $\ddot{\phi}$ through one period for the ONRT at zero forward speed in regular waves. Results are given for $\omega/\omega_{0}$=0.87, 0.94, and 1.05 to show the behavior around the natural frequency and to correspond with the CFD regular wave cases.}\label{liss}
\end{figure}

Figure \ref{result} shows that the proposed method greatly improves the predicted amplitudes of forced responses when compared to the linear benchmark, especially in regions near the natural frequency $\omega_{0}$. In general, the results are in fair agreement with the CFD cases. The captured shift in the natural frequency suggests the ML model imparts corrections to both the roll restoring and added mass moments, and the reduced amplitude near the natural frequency is indicative of corrections to the damping moment. The large shift may also be attributed to the effects of the appendages, which are present in the proposed model. The model still reproduces the long- and short-wave limits of the linear model, which indicates that the learned moments vanish in wave conditions which are dominantly linear.\par

However, the learned model over-predicts the amplitude of the response at wave excitation frequencies that are above the natural frequency, in a region where the mass terms of Eq. \eqref{nceom} dominate. It is important to note that the wave excitation model given by Eq. \eqref{wmom} makes a long-wave assumption, which could be partly responsible for the deviation, but the extent of this assumption is not known. To investigate this behavior further, Figure \ref{liss} shows the Lissajous curves for the predicted and CFD-derived roll moment correction $\delta$ against the predicted state $\phi$, $\dot{\phi}$, and $\ddot{\phi}$ for three cases around the natural frequency: $\omega/\omega_{0}$=0.87, 0.94, and 1.05. The progression of each curve is in the counter-clockwise direction. The curves are useful because they magnify any differences in phase and magnitude to make the underlying dynamics more apparent. One should keep in mind, however, that the CFD-derived moment also includes nonlinearity imparted by the incident waves and interacting radiated waves--a potentially major contribution which is not present in the learned model due to the unforced training data set.\par

In the $\omega/\omega_{0}$=0.94 wave case, which is the closest to the natural frequency of the three cases, both the magnitude and phase of the predicted moment correction are closest to the CFD-derived counterparts. It is also shown that the CFD-derived moments are primarily in phase with the position, and about 90$^\circ$ out of phase with the velocity. The spread of the predicted moments along the vertical axis indicates a negative (lagging) phase shift with respect to position. In the higher-frequency $\omega/\omega_{0}$=1.05 case, the ML model is considerably over-predicting the amplitude of the moment correction and a phase error is also present. This is likely the reason for the over-prediction in the response amplitude shown in Figure \ref{result}, as the phase angle may lead to an amplification of the response, instead of an attenuation. In the lower-frequency $\omega/\omega_{0}$=0.87 case, the amplitude of the predicted $\delta$ decreases significantly yet the phase agreement remains fair. Despite this, the small amplitude does not significantly affect the predictions as this region of the response is dominated by restoring forces, and the CFD-derived corrections show an almost purely linear form in phase with position.\par

Of interest in Figure \ref{liss} is the asymmetry in almost all of the curves, which differs strongly from classical linear-quadratic or linear-cubic models, especially in the higher-frequency case. The asymmetry may be partly attributed to nonlinear variation in the added mass, which is known to be asymmetric \citep{bigalabo2022}. The non-zero correction for the CFD-derived values at $\phi$=$\dot{\phi}$=$\ddot{\phi}$=0, especially in the higher-frequency case, is also attributable to a known mean-shift in the nonlinear responses in waves \citep{bigalabo2022}.\par

The comparison in Figure \ref{liss} exposes some weaknesses when training on only unforced responses such as roll decay time series, which is noted by other studies \citep{somayajula2017}, \citep{rodriguez2020}. Foremost is the trained models reduced ability to learn phase, as evidenced by the weak phase shift across the three frequencies. In addition, the over-prediction of the moment correction in the higher-frequencies suggests that while training at $\omega_{0}$ captures the damping moments at the most critical frequencies, the predictions could be improved by training in forced responses at two or more frequencies. However, this should not diminish the practicality of unforced roll decay time series as a training data set, especially when derived from experiments, given the considerable improvement of the RAO in Figure \ref{result}.

\section{Conclusions}\label{sect:concl}

In this work, the neural-corrector method of \cite{marlantes2022} is extended to model roll motions of an ONRT hull at model scale in regular waves. The hybrid data-driven method is trained using unforced roll decay time series data generated using CFD. Several conclusions can be drawn from the work:
\begin{itemize}
\item When trained on roll decay data, the proposed method learns a significant component of the nonlinear added mass, restoring, and damping moments. The effects of appendages, such as bilge keels, are also included.
\item Though the training data set is only unforced roll decay time series, the proposed method greatly improves predictions of forced responses in waves, with fair agreement compared to CFD validation data.
\item The proposed model reproduces the long- and short-wave limits of the linear model, indicating that the learned moments vanish in wave conditions which are dominantly linear.
\item The model is inexpensive to evaluate with a computational cost of approximately 5\% real-time.
\item The model is practical. The training data set consists of only a single, 20-second CFD roll decay simulation and linear coefficients computed using frequency-domain strip theory.
\end{itemize}
However, there are a  couple limitations which should be mentioned. First, the model, as proposed, is sensitive to the accuracy of the low-fidelity wave excitation model in Eq. \eqref{nceom}. To overcome this limitation, the force correction in \cite{marlantes2022} captures components of the nonlinear wave excitation force and the wave elevation is included as an additional input feature in the ML model. However, this also requires training data from roll responses in the presence of waves, which differs from the unforced roll decay data used in the present work. If forced responses were used for training data, time series from responses in two or more regular wave frequencies--or perhaps irregular waves, which encompass many frequencies--may also improve the predicted moments in Fig. \ref{liss}. Second, this study considered only 1-DOF roll motion, specifically due to its similarity to classical parameter identification methods, but should be extended to at least 3-DOF (heave, sway, roll) to capture the influence of coupling with other modes. However, free decay data is only available for modes of motion which have a restoring force. For modes such as sway, a forced approach would be required to generate the training data.\par

\printcredits

\bibliographystyle{cas-model2-names}

\bibliography{oerefs.bib}

\end{document}